\documentclass[pdflatex,sn-mathphys-num]{sn-jnl}% Math and Physical Sciences Numbered Reference Style
\usepackage{graphicx}%
\usepackage{multirow}%
\usepackage{amsmath,amssymb,amsfonts}%
\usepackage{amsthm}%
\usepackage{mathrsfs}%
\usepackage[title]{appendix}%
\usepackage{xcolor}%
\usepackage{textcomp}%
\usepackage{manyfoot}%
\usepackage{booktabs}%
\usepackage{algorithm}%
\usepackage{algorithmicx}%
\usepackage{algpseudocode}%
\usepackage{listings}%
\usepackage{graphicx} % Required for inserting images
\usepackage{amsmath}
\usepackage{booktabs}
\usepackage{amssymb}
\usepackage{amsthm}
\newtheorem{theorem}{Theorem}
\newtheorem{corollary}{Corollary}[theorem]
\newtheorem{lemma}[theorem]{Lemma}
\newtheorem{definition}{Definition}[section]
\newtheorem*{remark}{Remark}
\usepackage{appendix} % optional, for enhanced appendix formatting
\theoremstyle{thmstyleone}%
\theoremstyle{thmstyletwo}%

\theoremstyle{thmstylethree}%

\begin{document}

\title[Economic Security of Multiple Shared Security Protocols]{Economic Security of Multiple Shared Security Protocols}


\author[1]{\fnm{Abhimanyu} \sur{Nag}}

\author[2]{\fnm{Dhruv} \sur{Bodani}}
\equalcont{These authors contributed equally to this work.}

\author[2]{\fnm{Abhishek} \sur{Kumar}}
\equalcont{These authors contributed equally to this work.}

\affil[1]{\orgdiv{University of Alberta}, \country{Canada}}

\affil[2]{\orgdiv{Catalysis Labs}, \country{Singapore}}


\abstract{
As restaking protocols gain adoption across blockchain ecosystems, there is a need for Actively Validated Services (AVSs) to span multiple Shared Security Providers (SSPs). This leads to stake fragmentation which introduces new complications where an adversary may compromise an AVS by targeting its weakest SSP. In this paper, we formalize the Multiple SSP Problem and analyze two architectures : an isolated fragmented model called Model $\mathbb{M}$ and a shared unified model called Model $\mathbb{S}$, through a convex optimization and game-theoretic lens. We derive utility bounds, attack cost conditions and market equilibrium that describe protocol security for both models. Our results show that while Model $\mathbb{M}$ offers deployment flexibility, it inherits lowest-cost attack vulnerabilities, whereas Model $\mathbb{S}$ achieves tighter security guarantees through single validator sets and aggregated slashing logic. We conclude with future directions of work including an incentive-compatible stake rebalancing allocation in restaking ecosystems.}




\maketitle

\section{Introduction and Background}
\label{sec:intro}

Restaking protocols \cite{team2024eigenlayer} enable validators to reuse capital staked on base L1 blockchains \cite{nofer2017blockchain} (e.g., Ethereum \cite{buterin2013ethereum}, Solana \cite{yakovenko2018solana}, Bitcoin \cite{nakamoto2008bitcoin}) to secure external services, similar to asset rehypothecation \cite{monnet2011rehypothecation}. Platforms like Eigenlayer \cite{team2024eigenlayer}, Symbiotic \cite{symbiotic2025}, Babylon \cite{babylonlabs2025}, and Jito \cite{jitolabs2025} implement this via opt-in mechanisms, forming a modular architecture where \emph{Shared Security Providers} (SSPs) extend cryptoeconomic security to \emph{Actively Validated Services} or most recently \emph{Autonomous Verifiable Services} (AVSs). Each AVS may specify its own consensus, slashing, and incentive rules while inheriting security from the base chain’s validator set.

Each SSP operates as an economically independent domain including its native asset, price volatility and validator set and AVSs spanning multiple SSPs become fragmented. An attacker may corrupt global AVS safety by targeting the weakest constituent SSP. Under practical Byzantine Fault Tolerance (PBFT) \cite{castro1999practical} assumptions, a single local failure may compromise global liveness or safety.

There is a need to consolidate the fragmented economic security that these restaking protocols provide AVSs. Formalized as the "Multiple SSP Problem", this paper considers and compares two types of multiple SSP architectures that allows an AVS to borrow economic security of not just one but many L1 blockchains and provides a game-theoretic view of the risks and benefits of each type of architecture. Noticeably one of those architectures considers a fragmented stake model while the other considers a unified stake model. We survey some related work beforehand.

Until now, the EigenLayer whitepaper~\cite{team2024eigenlayer} conceptualizes restaking as reusable security for middleware, but does not quantify attack surfaces across SSPs. Durvasula and Roughgarden~\cite{durvasula2024robust} introduce $\gamma$-security, showing how adversarial spillover is bounded by $(1 + 1/\gamma)\psi$, where $\psi$ is the initial stake compromised. Chitra and Pai~\cite{chitra2024much} generalize this to strategic agents, proposing submodular reward mechanisms that align validator incentives with network-wide robustness. Our models, on the other hand, mimic classic results in portfolio theory~\cite{markowitz2010portfolio}, where stake allocations must be optimized under correlated risk and attack constraints. We take inspiration from the recent paper by Bar-Zur and Eyal \cite{bar2025elastic} for the modelling.

We develop a convex optimization \cite{boyd2004convex} and game-theoretic \cite{owen2013game} framework to evaluate validator security, attack cost and equilibrium behavior. Our contributions include:
\begin{enumerate}
    \item Definitions of weak and strong economic security in multi-SSP systems;
    \item A formal model of adversarial bribery feasibility and its relation to stake fragmentation;
    \item Security bounds and utility analysis under volatility and allocation heterogeneity;
    \item A maximin equilibrium characterization for validator restaking.
\end{enumerate}

We analytically derive conditions for validator-level safety, cost of attack and system-level robustness. Some simulation results are provided to support the theory, highlighting the tradeoffs between the two models.

\section{Problem Description}\label{probdesc}
We formally define a shared security protocol as follows:
\begin{definition}
    Shared Security Protocols (SSPs) are middleware entities that coordinate economic security from base blockchains to applications (AVSs) built on top.
\end{definition}

Notice that this is equivalent to the definition of \textit{Restaking Platforms} \cite{team2024eigenlayer}.
With the increasing number of SSPs, our main problem statement can be formalized as follows :

\begin{quote}
\textbf{\textit{How can an AVS inherit security from multiple heterogeneous SSPs, and what are the associated guarantees and risks?}}
\end{quote}

Assuming the AVS uses a PBFT-style protocol as in \cite{castro1999practical} with $n$ nodes (up to $f$ Byzantine), authenticated and eventually synchronous communication, and a designated leader node per consensus instance, we study two designs:

\begin{itemize}
    \item \textbf{Isolated Multi-SSP Model ($\mathbb{M}$)}: AVS instances run separately on each SSP, with independent consensus, incentives, and slashing.
    \item \textbf{Shared Multi-SSP Model ($\mathbb{S}$)}: An AVS spans all SSPs with shared logic, but validators restake variably across domains.
\end{itemize}

The motivation for this study is to evaluate the economic security of aggregator protocols like Catalysis \cite{catalysis2025} and compare the increased economic security of the design to solve this problem, model $\mathbb{S}$ adopted by Catalysis, versus the alternative model $\mathbb{M}$. 

\begin{figure}
    \includegraphics[width=0.5\linewidth]{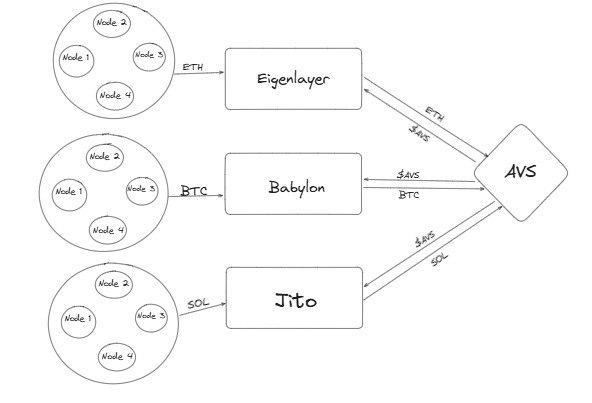}
    \includegraphics[width=0.5\linewidth]{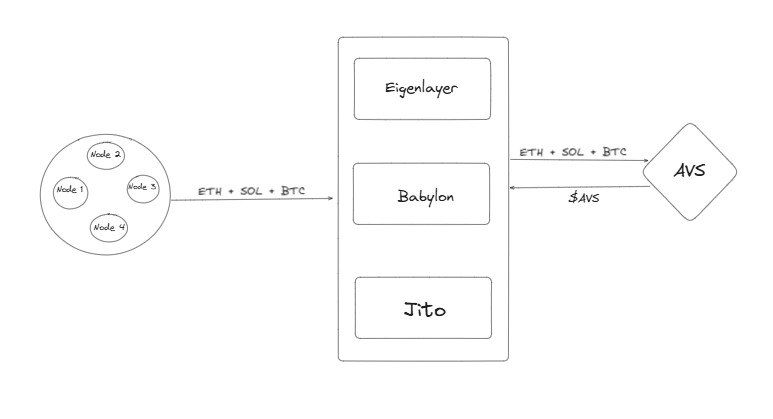}
    \caption{Model $\mathbb{M}$ (left) with isolated fragmented restaking primitives while Catalysis Model $\mathbb{S}$ (right) with a single consensus pool with a restaking abstraction layer}
    \label{models}
\end{figure}

We depart from prior network topological study by analyzing restaking as an economic problem across multiple SSPs, each as cost-of-corruption frameworks, inspired by StakeSure~\cite{deb2024stakesure}. Building on the formalism of Bar-Zur and Eyal ~\cite{bar2025elastic}, we model the security guarantees and simulate the economic security of both models.
Figure \ref{models} shows a graphical representation of the two designs in consideration.

We see that while $\mathbb{M}$ offers flexibility, it suffers from stake fragmentation, greater operator workload, lower attack costs and higher attack surfaces. $\mathbb{S}$ achieves tighter security bounds under reasonable assumptions.

\section{Model}
\subsection{Setup}
A table of notations can be found in Appendix \ref{notations}.\\
For simplicity, let there be a single AVS with a deterministic profit from corruption $\pi$ and attack threshold $\theta$ (i.e if the operators capture stake proportion above $\theta$, usually $\frac{1}{3}$ in PBFT, an attack would be successful) and reward $R$ (considered in stablecoin \$USDC for the purposes of this with paper) and APY $r$\% of total stake.
Consider $n$ node operators denoted as set $V = \{v_1,v_2,v_3, \cdots, v_n\}$, a set of $k$ SSPs denoted by $S = \{s_1,s_2,s_3, \cdots, s_k\}$.  Borrowing model notations from \cite{bar2025elastic}, we formally define $\sigma : V \to \mathbb{R}_{>0}$ and $\omega : V \times S \to \mathbb{R}_{>0}$ such that each node operator has total stake $\sigma(V) \in \mathbb{R}_{>0}$ with allocation for each SSP being $\omega(V,S) \in \mathbb{R}_{>0}$. Given each SSP has a unique asset which is restaked through it (such as ETH,SOL,BTC etc) and 
$$\sum_{i = 1}^{k} \omega(v_j,s_i) = \sigma(v_j)$$ for any node operator $v_j \in V$.\\
The total economic security enjoyed by an AVS is equivalent to 
$$\sum_{j = 1}^{n}\sum_{i = 1}^{k} \omega(v_j,s_i) = \sum_{j = 1}^{n}\sigma(v_j) = \Delta$$ and the total economic security enjoyed by the $j$th SSP pool is 
$$\sum_{i = 1}^{n} \omega(v_i,s_j) = \Delta_j$$
where $\Delta$ is used to denote the total stake in all SSPs combined while $\Delta_j$ is used to denote the total stake in $j$th SSP pool.\\
Let the feasible allocation set be:
\[
\Omega := \left\{ \omega \in \mathbb{R}^{n \times k}_{\geq 0} \,:\, \sum_{j=1}^k \omega(v_i, s_j) = \sigma(v_i), \, \forall i \right\}
\]

% \begin{remark}
%     We will sometimes use $p_j(t)$ to represent USD value of asset used in SSP $j$ to talk about price volatilites later in this section. Explicit price analysis of the asset for the SSP is out of scope for this paper.
% \end{remark}
 For simulations, we take $20$ node operators across $10$ SSPs where we ran a random sample of stake values from a Uniform distribution ranging from $10$ to $100$ units and allocations are randomly sampled from a Dirichlet distribution \cite{lin2016dirichlet}. The rewards were uniformly sampled from a Uniform distribution from $5$ to $20$ units (equivalent to a uniform distribution of $r$ from $0.05$ to $0.5$). We ran 1000 monte carlo simulations \cite{james1980monte} over a constant sampling of $\pi$ from Uniform distribution from $(10000,80000)$. For each asset we take latest USD price as the dollar denominated price during simulation. We include a discussion about our choices about the values of $\pi,R, \theta$ and $r$ and their relevance in Section \ref{discussion}.
\subsubsection{Shared Security Protocol Models Formal Description}
We formally distinguish between the two classes of Multiple SSP architectures relevant to our analysis:

\begin{itemize}
    \item \textbf{Model }$\mathbb{M}$ : Let $\mathbb{M} := \{P_1, P_2, \dots, P_k\}$ denote a collection of $k$ BFT consensus pools, each associated with a distinct SSP. The total stake $\Delta$ is partitioned across these pools such that
    \[
    \sum_{j=1}^{k} \Delta_j = \Delta, \quad \text{where } \Delta_j \text{ is the stake allocated to pool } P_j.
    \]
    Each pool $P_j$ operates independently, slashing conditions and economic dynamics. Validators may exhibit heterogeneous behavior across pools, implying differentiated failure and bribery thresholds for each SSP.

    \item \textbf{Model }$\mathbb{S}$ : In contrast, $\mathbb{S}$ consists of a single Byzantine Fault Tolerant consensus pool securing all $k$ SSPs simultaneously. The stake $\Delta$ is partitioned into allocations $\{\Delta_1, \Delta_2, \dots, \Delta_k\}$ across SSPs, but the validator set is unified, and all participants are subject to the same consensus rules and slashing mechanism. While economic exposure varies by service, operational behavior remains consistent across SSPs.

    % \item \textbf{Model }$\mathbb{S}_{\text{single}}$ : Single-SSP model (e.g., Eigenlayer alone) is the degenerate case, in which only one SSP is active. Such a system has a narrower volatility and risk profile. 
\end{itemize}
% Therefore,
% \[
% \Delta_{\text{single}} \ll \Delta_{\mathbb{S}},\ \Delta_{\mathbb{M}}.
% \]
\subsubsection{Utility Analysis for Node Operators}
Now to help with developing our utility model,  let $\alpha : V \times S  \to \mathbb{R}_{>0}, \alpha(V,S) \in [0,\omega(V,S)]$ be the amount of allocated stake in an SSP that can be used to attack the AVS by a node operator, all other things remaining constant. \\
The cost to each node $v$ to attack that AVS can be written as 
$$c(v,\alpha) = min(\sigma(v),\sum_{s \in S} \alpha(v,s))$$
Hence the total cost to attack the AVS is 
$$C(\alpha) = \sum_{v \in V} c(v,\alpha)$$
Now coming back to the AVS, given attack threshold $\theta$, attacking this AVS will be profitable per SSP iff $\frac{\sum_{v \in V} \alpha(v,s)}{\sum_{v \in V} \omega(v,s)} \geq \theta$

If a successful attack happens on the AVS, the distribution of the attack prize will be ideally proportional to the costs borne by the node operators to conduct the attack. i.e 
\[
\gamma(v,\alpha) =
\begin{cases}
\frac{c(v,\alpha)}{C(\alpha)} & C(\alpha) > 0 \\
\end{cases}
\]
\textbf{Case 1: A successful attack occurs}\\
When an attack is successful and assuming costs are incurred, we derive the utility of a node operator $v$ who decides to contribute stake $\alpha$ to attack the AVS as follows :
$$u_{v}(\alpha)= \gamma(v,\alpha)\cdot \pi - c(v,\alpha)
    = \sum_{v} \alpha(v,s) (\frac{\pi}{\sum_{s} \sum_{v} \alpha(v,s)} - 1)$$
This implies that an attack will be profitable if and only if 
$$\theta \cdot \Delta <\sum_{s} \sum_{v} \alpha(v,s) < \pi$$
\textbf{Case 2 : When there is no attack}\\
When there is no attack, keeping everything constant, we consider the utility of the node operator solely depends on the rewards from the AVS. Let us assume an APY based reward system where rewards are emitted at a rate of $r$\% of the total stake $\Delta$. Using a proportional reward system where rewards corresponding to SSP $i$ is $R_i$ thus total rewards $\sum_{i = 1}^{k} R_i = r\cdot\Delta = R$. There is a proportional distribution of rewards per SSP (as the best way to maintain economic equality in the protocol, see Appendix \ref{rewardapp} for a detailed discussion and full proof of utility) 
which implies rewards for any node operator $v_i$ is 
$$ u(v_i) = \frac{R}{\Delta} \cdot \sigma(v_i) = r\cdot \sigma(v_i)$$

\subsection{Security Guarantees and Main Results}
We outline some main results in this section that pertain to the multiple SSP model ($\mathbb{S}$ or $\mathbb{M}$ or both). Full proofs can be found in Appendix \ref{lemproof}.

A protocol is said to be economically secure if attacking it costs more than the profit from the attack \cite{deb2024stakesure}. That is, the attack cost must be greater than or equal to the attack payoff. Using this we make a few claims about both the models,
\begin{definition}[Weak Shared Security] \label{WES}
    An AVS is cryptoeconomically secure in a shared security primitive if and only if
    \[
    \frac{\pi}{\Delta} \leq \theta.
    \]
\end{definition}

We include a stronger definition about shared security with a few corresponding lemmas. The intuition is that for stronger incentive to be honest the average utility across all validators from being honest has to be greater than that of being dishonest. Therefore,

\begin{definition}[Strong Shared Security] \label{SES}
An AVS is said to be strongly secure in a shared security primitive if:
\[
\frac{1}{n} \sum_{i=1}^{n}\sum_{j = 1}^{k} \alpha(v_i,s_j) + \frac{R}{n} > \pi.
\]
\end{definition}

\begin{lemma}[Validator Security Bound]
\label{lemma:validator-margin}
The average validator utility from honest participation has an upper bound:
\[
\frac{(\theta + r)\cdot \Delta}{n} > \frac{1}{n} \left( \sum_{j=1}^{k} \sum_{i = 1}^{n} \alpha(v_i,s_j) + r \cdot \Delta \right).
\]
\end{lemma}

Combining definition \ref{SES} and lemma \ref{lemma:validator-margin} we have
\[
\frac{(\theta + r)\cdot \Delta}{n} > \frac{1}{n} \left( \sum_{j=1}^{k} \sum_{i = 1}^{n} \alpha(v_i,s_j) + r \cdot \Delta \right) > \pi
\]

which obviously implies that to satisfy security guarantee of the multiple SSPs, the number of validators must obey:
\[
n < \frac{(\theta +r)\cdot \Delta}{\pi}.
\]
which implies at high cost of corruption and low stake value, the system should reorganise towards centralization with a lower number of SSPs as well as a lower number of node operators per SSP and decentralization otherwise  in case of higher APY and security threshold with a greater amount of nodes.
We now shift our focus to deriving some specific properties between the $\mathbb{S}$ and $\mathbb{M}$ models of multiple SSPs.
\begin{lemma}\label{attsur}
In a multi-SSP model $\mathbb{M}$, the attack surface increases compared to the single-SSP model. It remains constant in $\mathbb{S}$. Further, the minimum cost to attack an AVS is lower in model $\mathbb{M}$ than in $\mathbb{S}$. 
\end{lemma}

\begin{figure}
    \centering
    \includegraphics[width=0.5\linewidth]{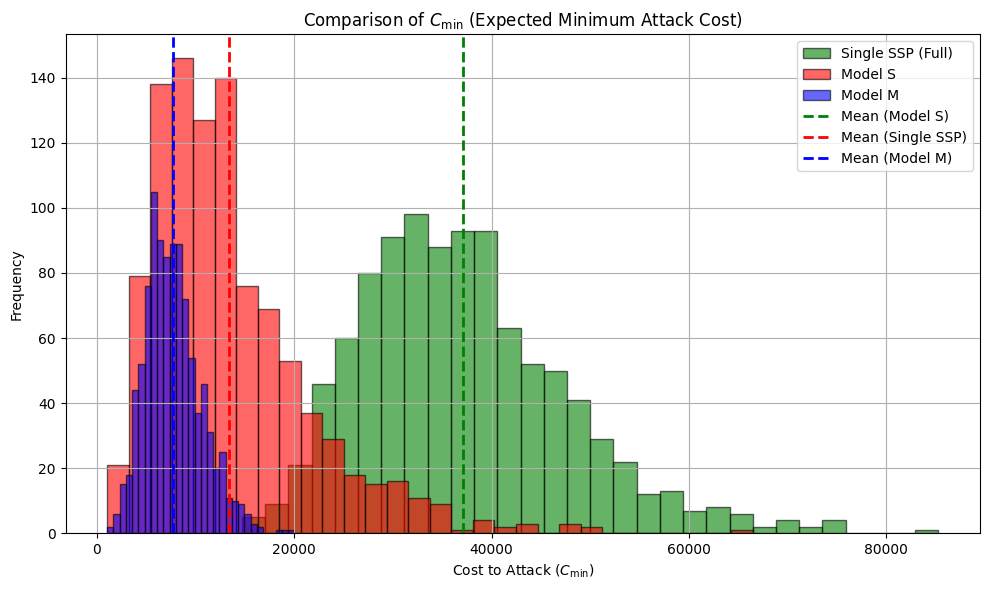}
    \caption{A comparison of minimum cost to attack the AVS in the three models. Cost to attack model $\mathbb{S}$ $\geq$ Cost to attack a single SSP $\geq$ Cost to attack model $\mathbb{M}$}.
    \label{costattack}
\end{figure}

Figure \ref{costattack} shows a visual representation of the comparison of cost of attacking the AVS in Model $\mathbb{S}$, Model $\mathbb{M}$ as well as a single SSP.

This motivates the following corollary:

\begin{corollary}[Security Threshold Tightening]
In $\mathbb{M}$, security requires:
\[
\pi < \theta \cdot \min_j \Delta_j,
\]
instead of $\pi < \theta \cdot \Delta$.
\end{corollary}

Thus, model $\mathbb{M}$ lowers the system-wide security threshold while $\mathbb{S}$ actually increases it because of increased stake control. 

To add to this, when we incorporate price behavior and the volatility of the underlying asset, we can make a claim that is reminiscent of optimal portfolio risk theory \cite{markowitz2010portfolio}.

Let $\Delta_j(t) = \sum_{v \in V} \omega(v, s_j) \cdot p_j(t)$, where $p_j(t)$ is the USD price of the underlying asset, be the dollar-denominated security of SSP \( s_j \). Then:
\[
\operatorname{Var}[\Delta_j(t)] = \sum_{v} \omega(v, s_j)^2 \cdot \operatorname{Var}[p_j(t)] + \sum_{v \neq v'} \omega(v, s_j) \omega(v', s_j) \cdot \operatorname{Cov}[p_j(t), p_j(t)].
\]

Suppose two SSPs, $s_j$ and $s_k$, restake tokens with price processes $p_j(t)$ and $p_k(t)$ that are uncorrelated, i.e.,
\[
\operatorname{Cov}[p_j(t), p_k(t)] = 0.
\]
Then, for a validator $v$ that restakes in both SSPs, the joint volatility of their total security contribution
\[
\Delta_v(t) = \omega(v, s_j) \cdot p_j(t) + \omega(v, s_k) \cdot p_k(t)
\]
is minimized (in expectation) compared to the correlated case.

This demonstrates that using economically uncorrelated tokens for restaking reduces volatility in validator-level economic security.

\begin{remark}
In practice, crypto assets are highly correlated during extreme events due to shared market sentiment, regulatory shocks, or liquidity-driven drawdowns. As such, while uncorrelated restaking assets may reduce day-to-day volatility, they may offer limited protection against Black Swan events or correlated slashing cascades in model $\mathbb{M}$. See \cite{nie2022analysis}.
\end{remark}

As we see in Table \ref{tab:correlation_matrix}, we compare the correlation matrix among the three underlying assets in question : ETH,BTC and SOL.
\begin{table}[h]
\centering
\caption{Empirical correlation matrix of daily returns among staking assets (BTC, ETH, SOL). All values computed using log returns over a one-year window.}
\label{tab:correlation_matrix}
\begin{tabular}{lccc}
\toprule
\textbf{Asset} & \textbf{BTC} & \textbf{ETH} & \textbf{SOL} \\
\midrule
BTC & 1.0000 & 0.8194 & 0.7631 \\
ETH & 0.8194 & 1.0000 & 0.7307 \\
SOL & 0.7631 & 0.7307 & 1.0000 \\
\bottomrule
\end{tabular}
\end{table}

\begin{figure}
    \centering
    \includegraphics[width=0.5\linewidth]{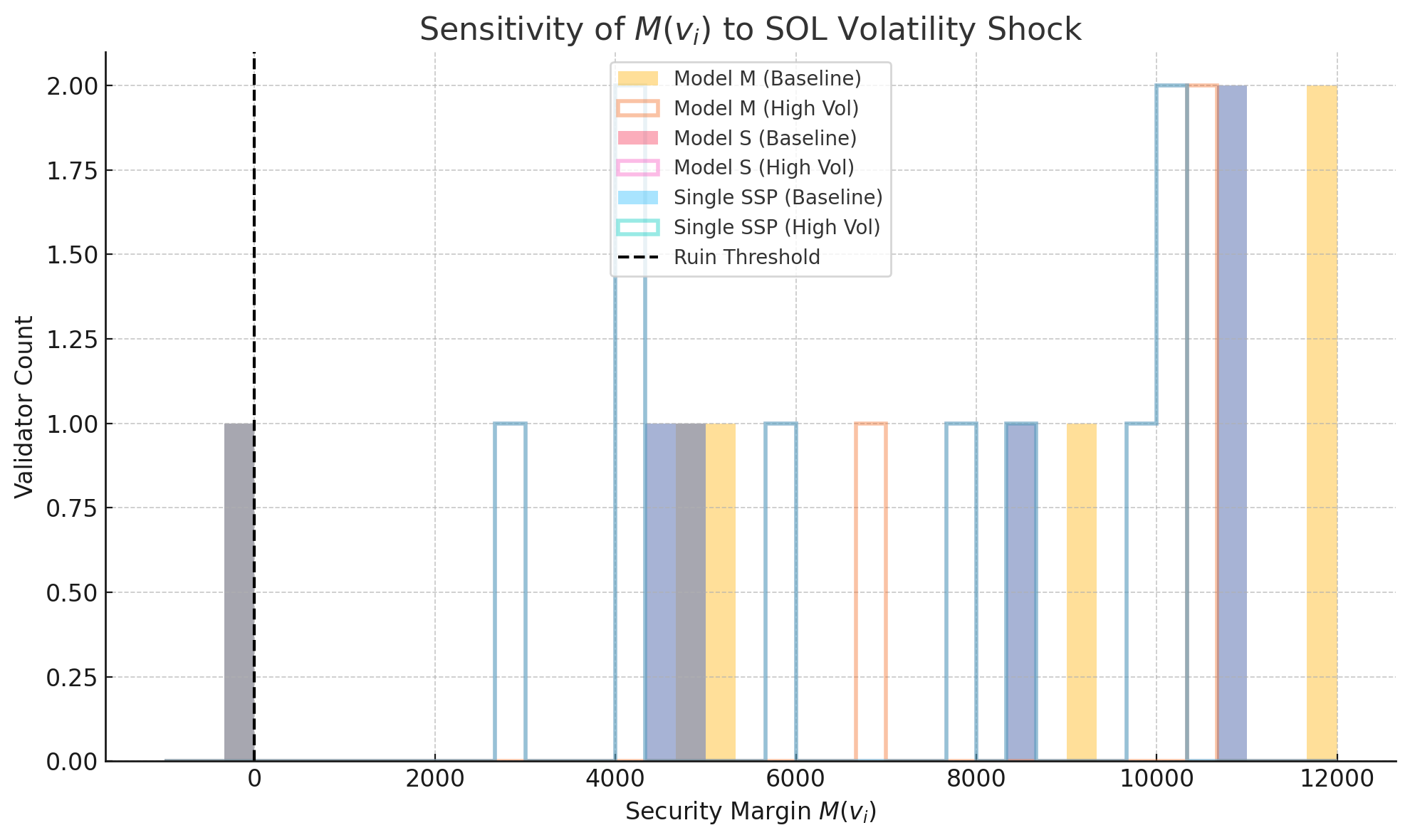}
    \caption{Histogram of validator utility from honest participation ($M(v_i)$) across different models under baseline and high-volatility conditions. Under a SOL volatility shock, Model $\mathbb{M}$ shows a clear shift toward lower security margin, while Model $\mathbb{S}$ and Single SSP remain comparatively resilient. The dashed line indicates the ruin threshold ($M(v_i)<0$).}
    \label{sensitivity}
\end{figure}

The high pairwise correlations indicate that the three assets respond similarly to macroeconomic shocks which would diminish diversification across SSPs. Therefore,

\begin{remark}
    In order to decrease the risk of correlated slashing or cascading failures, there is a need to use slightly correlated or uncorrelated assets as securities in the two models. While it should not affect Model $\mathbb{S}$ as adversely as stake distribution is not fragmented in Model $\mathbb{S}$ unlike that in $\mathbb{M}$, it would still decrease overall security for both models. See figure \ref{sensitivity}
\end{remark}

Also in model $\mathbb{M}$, if even one SSP is consistently weaker than another (in a probabilistic sense), then the system’s minimum security is dragged down by it no matter how strong the others are. In expectation, the attacker’s cost is capped by the weakest acceptable SSP. That is,
% \begin{proof}
% If $\omega(v, s_j)$ is deterministic and $p_j(t)$ is random, then:
% \[
% \Delta_j(t) = p_j(t) \cdot \sum_v \omega(v, s_j) = p_j(t) \cdot W_j.
% \]
% Thus:
% \[
% \operatorname{Var}[\Delta_j(t)] = W_j^2 \cdot \operatorname{Var}[p_j(t)].
% \]
% If both $\omega(v, s_j)(t)$ and $p_j(t)$ are random:
% \[
% \operatorname{Var}[\Delta_j(t)] = \sum_v \operatorname{Var}[\omega(v, s_j)] \cdot \mathbb{E}[p_j(t)]^2 + \sum_v \mathbb{E}[\omega(v, s_j)]^2 \cdot \operatorname{Var}[p_j(t)] + \text{cross-terms}.
% \]
% \end{proof}
\begin{lemma}[Stochastic Dominance \cite{levy1992stochastic} and Security Bounds] \label{stochdom}
Let \( \{\Delta_j(t)\}_{j=1}^k \) be non-negative random variables representing security levels of $k$ SSPs, and let \( C_{\min} = \theta \cdot \min_j \Delta_j(t) \). If for some $j,k$:
\[
\Delta_j(t) \prec_{FSD} \Delta_k(t),
\]
then:
\[
\mathbb{E}[C_{\min}] \leq \theta \cdot \mathbb{E}[\Delta_k(t)].
\]
\end{lemma}
\begin{figure}
    \centering
    \includegraphics[width=0.5\linewidth]{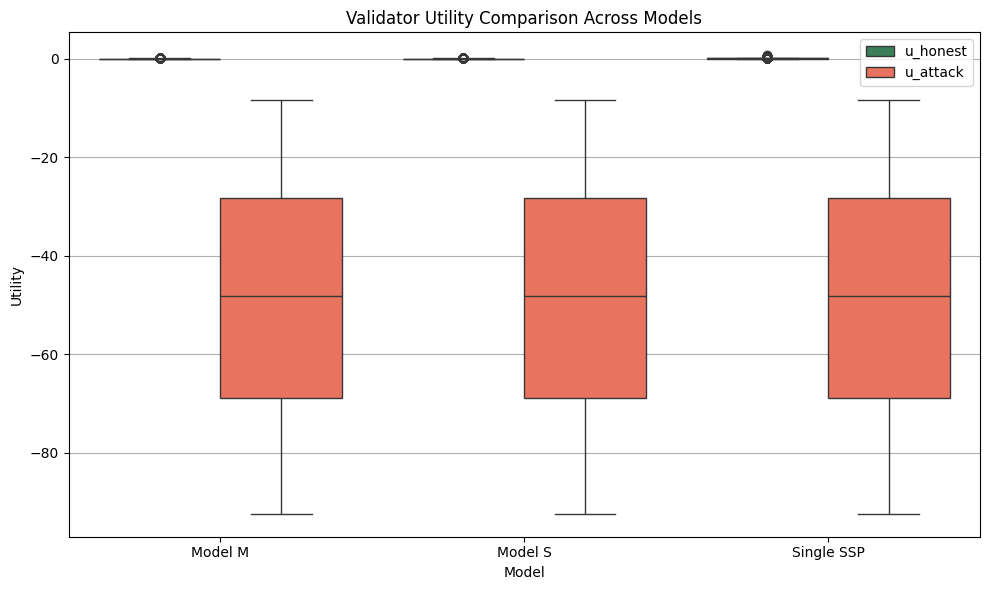}
    \includegraphics[width=0.5\linewidth]{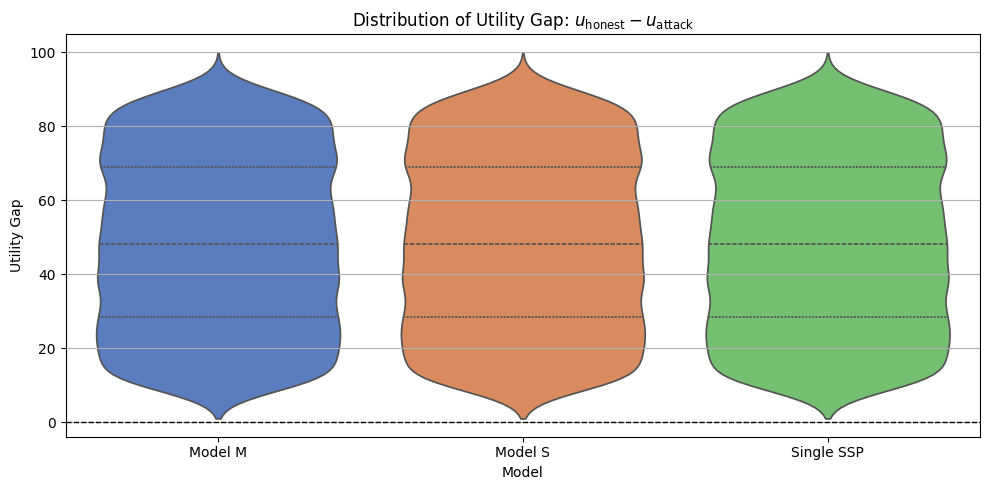}
    \caption{side-by-side comparison for validator utility from being honest and utility from attacking}
    \label{fig:utilattack}
\end{figure}
% \begin{proof}
% FSD implies:
% \[
% \mathbb{P}(\Delta_j(t) \leq x) \geq \mathbb{P}(\Delta_k(t) \leq x), \quad \forall x.
% \]
% Then for monotonic increasing $f$:
% \[
% \mathbb{E}[f(\Delta_j(t))] \leq \mathbb{E}[f(\Delta_k(t))].
% \]
% Taking $f(x) = x$:
% \[
% \mathbb{E}[\Delta_j(t)] \leq \mathbb{E}[\Delta_k(t)].
% \]
% Since:
% \[
% \min_j \Delta_j(t) \leq \Delta_k(t),
% \quad \Rightarrow \mathbb{E}[\min_j \Delta_j(t)] \leq \mathbb{E}[\Delta_k(t)].
% \]
% Multiplying both sides by \( \theta \):
% \[
% \mathbb{E}[C_{\min}] \leq \theta \cdot \mathbb{E}[\Delta_k(t)].
% \]
% \end{proof}

We finally get to what arguably is the most important result of all :
\begin{theorem}[Optimal Allocation Minimizes Minimum Risk]
\label{lem:concavity}
Let \( x = (x_1, \dots, x_k) \in \mathbb{R}_{\geq 0}^k \) denote the security levels of the \(k\) SSPs in the multiple SSP model \( \mathbb{M} \), i.e., \( x_j = \Delta_j \). Then the cryptoeconomic security level of the AVS, defined by the minimum cost to corrupt the weakest SSP,
\[
S(x) := \theta \cdot \min_j x_j,
\]
is a concave function over \( \mathbb{R}_{\geq 0}^k \).

Consider the optimization problem of maximizing the minimum  security across all SSPs:
\[
\max_{\omega \in \mathbb{R}_+^{n \times k}} \min_{j \in [k]} \left( \sum_{i=1}^n \omega(v_i, s_j) \right),
\]
subject to the per-validator stake conservation constraints:
\[
\sum_{j=1}^k \omega(v_i, s_j) = \sigma(v_i), \quad \forall i \in [n],
\]
\[
\omega(v_i, s_j) \geq 0, \quad \forall i \in [n], \; j \in [k].
\]
Then, the optimal solution is attained when all \( \Delta_j := \sum_{i=1}^n \omega(v_i, s_j) \) are equal.
\end{theorem}

Theorem~\ref{lem:concavity} shows that under adversarial cost minimization, the optimal allocation of stake \(\omega^*\) from the set of all feasible validator-to-SSP stake assignments must equalize the security levels \(\Delta_j\) across all SSPs to maximize security of model $\mathbb{M}$ while no such balance is needed for security of Model $\mathbb{S}$ (note that there is no incosistency with our definitions for equilibrium \ref{Equil}). This follows from \cite{boyd2004convex}.

Formally, this means that:
\[
\omega^* \in \arg\max_{\omega \in \Omega} \min_j \Delta_j(\omega) \quad \Rightarrow \quad \Delta_j^* = \Delta_\ell^*, \quad \forall j, \ell \in [k].
\]

Hence, the concavity of the minimum security function combined with this optimality condition implies that Model \(\mathbb{M}\) is economically robust \emph{if and only if} the validator stake is symmetrically distributed to neutralize weakest-link failure. This constraint not required in Model \(\mathbb{S}\), which inherits a more centralized failure mode but with higher aggregate stake per AVS which makes it a safer alternative.

\subsection{Market Equilibrium}
We now formalize the notion of equilibrium in a restaking market across multiple SSPs, wherein validator behavior aligns with maximal economic security.

\begin{definition}[Market Equilibrium]
A multiple-SSP restaking configuration $\omega^* \in \Omega$ is said to be a \emph{market equilibrium} if:
\begin{enumerate}
    \item Each validator $v_i$ maximizes their utility $u(v_i)$ given induced SSP security levels $\{\Delta_j\}_{j=1}^k$,
    \item The minimum cost of attack is maximized over all feasible allocations $\omega \in \Omega$,
    \item No validator has a unilateral incentive to deviate from their assigned allocation in $\omega^*$.
\end{enumerate}
\end{definition}

This leads to the following maximin convex optimization formulation:
\[
\max_{\omega \in \Omega} \min_j \Delta_j \quad \text{s.t.} \quad u(v_i; \omega) \geq u^*, \quad \forall i.
\]

This program seeks to maximize the weakest security level across SSPs while preserving individual incentive compatibility. Since $\Delta_j$ is linear in $\omega$ and $u(v_i)$ is convex in stake allocation under proportional reward schemes, the problem is solved using convex programming \cite{boyd2004convex}.

\begin{lemma}[Optimal Allocation Equalizes Security]
\label{lem:equal-delta}
At the optimal solution $\omega^*$ to the above problem, the security levels are equalized across SSPs:
\[
\Delta_j = \Delta_\ell \quad \forall j, \ell \in \{1, \dots, k\}.
\]

At equilibrium, the minimum cost of attack satisfies:
\[
C_{\min}^* = \theta \cdot \min_j \Delta_j = \theta \cdot \frac{1}{k} \sum_{j=1}^k \Delta_j.
\]
\end{lemma}

This shows that equalizing $\Delta_j$ not only satisfies optimality but also maximizes the security threshold uniformly in both models.

\begin{lemma}[Equilibrium under Proportional Rewards]
\label{Equil}
Suppose the AVS distributes rewards proportionally to stake. If all SSPs have equal total stake (i.e., $\Delta_j = \Delta/k$ for all $j$), then validator utility $u(v_i)$ is independent of allocation, and equilibrium allocation $\omega^*$ maximizes both utility and security.
\end{lemma}

\begin{remark}
    In both Model $\mathbb{S}$ and Model $\mathbb{M}$, an equal-$\Delta_j$ allocation is a Nash equilibrium \cite{kreps1989nash} and globally optimal for security. Validators face no incentive to reallocate, and the protocol achieves maximum resistance to the lowest-cost attack. In every other case Model $\mathbb{S}$ has the most robust security between the two. The optimal security level can be reached for Model $\mathbb{M}$ (as well as Model $\mathbb{S}$) using stake rebalancing methods across SSPs.
\end{remark}

\section{Bribery Attack}

We add a small writeup on bribery attack thresholds. Let us consider a rational adversary capable of: (i) bribing validators, and (iii) targeting economically weak SSPs. Validators behave rationally, weighing honest rewards \(u(v_i)\) against bribes adjusted for slashing risk. Attacks are incentive-compatible and succeed only when protocol incentives misalign.

\subsection{Bribery-Based Attack Cost Comparison}

Each validator \(v_i\), that will get slashed equal to sl$(v_i)$, will accept a bribe \(b(v_i, s_j)\) only if:
\[
b(v_i, s_j) - \alpha(v_i, s_j) - \text{sl}(v_i) > u(v_i),
\]
where \(\alpha(v_i, s_j)\) is the portion restaked maliciously. Define the per-unit bribery requirement as \(\lambda_j(v_i) := \frac{u(v_i)}{\omega(v_i, s_j)}\). For a subset \(V^*_j \subseteq V\) compromising \(s_j\), the cost becomes:
\[
C_j = \min_{V^*_j} \left\{ (1 + \lambda^*_j) \cdot \sum_{v \in V^*_j} \omega(v, s_j) \right\}, \quad \lambda^*_j = \max_{v \in V^*_j} \lambda_j(v).
\]
Total system cost for the multi-SSP model (Model $\mathbb{M}$):
\[
C_{\text{multi}} = \min_j C_j.
\]
For model $\mathbb{S}$:
\[
C_{\text{single}} = (1 + \lambda^*) \cdot \theta \cdot \Delta, \quad \lambda^* = \max_v \lambda(v), \quad \Delta = \sum_v \sigma(v) \cdot p.
\]

\begin{figure}
    \centering
    \includegraphics[width=0.5\linewidth]{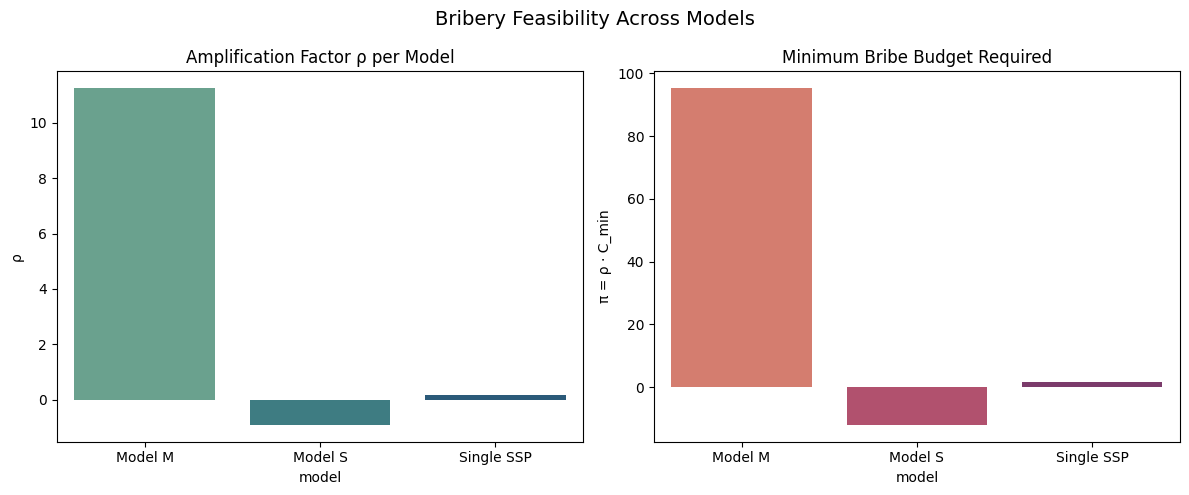}
    \caption{Model $\mathbb{M}$ has a strictly lower bribery threshold than Model $\mathbb{S}$ and is more susceptible to targeted attacks. Model $\mathbb{M}$ has significantly higher amplification factor but greater cost requirements. Models $\mathbb{S}$ and Single SSP demonstrate greater resistance with negative  amplification factors and bribery budgets for positive utility}
    \label{fig:bribery}
\end{figure}

\noindent
Model $\mathbb{M}$ typically lowers the attacker’s cost by allowing selection of the weakest target SSP, potentially with a low \(\lambda^*_j\) and small validator subset. In contrast, Model $\mathbb{S}$ concentrates all stake and rewards, making any bribery attack costlier due to higher coordination requirements and uniformly elevated \(\lambda^*\). This is shown in figure \ref{fig:bribery}. More can be found in \cite{sun2020model}.

\section{Conclusion, Discussion and Future Work} \label{discussion}

This work establishes a mathematically rigorous framework for evaluating security guarantees in multi-SSP restaking systems. We compare isolated (Model $\mathbb{M}$) and shared (Model $\mathbb{S}$) security primitives by modeling validator utility guarantees under adversarial presence and bribery feasibility. We prove that while Model $\mathbb{M}$ permits flexibility and modular deployments, it introduces convex risk profiles due to its lowest-cost attack vector. In contrast, Model $\mathbb{S}$ offers globally tighter security bounds due to its aggregate stake pool and unified slashing incentives.

There is a promising direction for future work: the design of incentive-compatible stake rebalancing protocols that converge to these equilibria under bounded rationality and limited observability. \emph{Stake Rebalancing} refers to the redistribution of staked assets across multiple SSPs to maintain system-wide economic security. This may mean movement of assets from say Eigenlayer to Symbiotic and so on.
In future work we will investigate how rebalancing can be achieved. Some ways include:
\begin{itemize}
    \item \textbf{Incentive Gradients:} SSPs have dynamic risk-reward scores \( R_i(t) \) that guide stakers to optimize expected returns. This should lead to self organization of stake.
    \item \textbf{Auction Mechanisms:} A new character called \emph{Coordinators} can periodically reallocate stake via sealed-bid auctions \cite{alvarez2020comprehensive}.
    \item \textbf{Stochastic Flow Limits:} Stake inflows/outflows can be constrained by protocol governed by coupled differential equations \cite{adomian1985coupled}.
\end{itemize}

At equilibrium, the marginal utility of stake allocation is equalized across SSPs:
\[
\frac{\partial U(s_j)}{\partial \Delta_j} = \lambda \quad \forall j \in {1,2,\cdots,k}
\]
where \( U(s_j) \) includes both reward and security-adjusted risk per SSP. This can be framed as another convex optimization \cite{boyd2004convex} problem under constraints.

Future research will also incorporate asset dynamics over time similar to what we see in standard crypto risk management literature \cite{gauntlet_compound2020},\cite{ghosh2024compoundv3economicaudit} to model black swan events and stress test the protocols. Also there is a case for considering stochastic stake reallocation, correlated validator preferences, and real-world validator behavioral models to further refine these security guarantees. Our framework also generalizes to heterogeneous asset collateralization and nested slashing domains. There is also a case for using tools like MAIDs \cite{nag2024multiagentinfluencediagrams} to assess incentive attack feasibility for protocol designers.

For our simulations, we sample the adversarial profit $\pi$, base reward rate $r$, and total AVS reward $R$ from independent uniform distributions over bounded intervals calibrated for qualitative analysis. We fix the protocol threshold parameter at $\theta = \frac{1}{3}$, consistent with typical PBFT-style fault tolerance assumptions. Given the current nascency of deployed AVSs and limited publicly available data from SSP platforms, precise estimation of these parameters remains challenging.
We adopt uniform sampling as a model-agnostic stress-testing approach, allowing us to evaluate system behavior across a wide range of plausible economic regimes without overfitting to a specific AVS design.

Future work may refine these assumptions using empirical distributions inferred from deployed systems. In particular, the development of EigenLayer’s “sidecar” modules \cite{eigenlayerSidecar2025},intended to report validator-specific restaking activity and slashing exposure, could provide  data on real-world reward rates, stake allocations, and AVS-specific bribery thresholds. Incorporating this into our model would allow for calibrated simulations and risk-aware parameter estimation in production-grade deployments.

From a system design perspective, our results imply Model $\mathbb{S}$ provides stronger cryptoeconomic security. Nevertheless, Model $\mathbb{M}$ can be secured under a maximin allocation regime provided validators align on equalizing $\Delta_j$ across SSPs. In the absence of incentive compatible stake rebalancing methods for now, Catalysis model $\mathbb{S}$ remains the strongest Multiple SSP architecture when it comes to the highest economic security provided to any AVS.

\bibliography{sn-bibliography}


\begin{thebibliography}{31}
\ifx \bisbn   \undefined \def \bisbn  #1{ISBN #1}\fi
\ifx \binits  \undefined \def \binits#1{#1}\fi
\ifx \bauthor  \undefined \def \bauthor#1{#1}\fi
\ifx \batitle  \undefined \def \batitle#1{#1}\fi
\ifx \bjtitle  \undefined \def \bjtitle#1{#1}\fi
\ifx \bvolume  \undefined \def \bvolume#1{\textbf{#1}}\fi
\ifx \byear  \undefined \def \byear#1{#1}\fi
\ifx \bissue  \undefined \def \bissue#1{#1}\fi
\ifx \bfpage  \undefined \def \bfpage#1{#1}\fi
\ifx \blpage  \undefined \def \blpage #1{#1}\fi
\ifx \burl  \undefined \def \burl#1{\textsf{#1}}\fi
\ifx \doiurl  \undefined \def \doiurl#1{\url{https://doi.org/#1}}\fi
\ifx \betal  \undefined \def \betal{\textit{et al.}}\fi
\ifx \binstitute  \undefined \def \binstitute#1{#1}\fi
\ifx \binstitutionaled  \undefined \def \binstitutionaled#1{#1}\fi
\ifx \bctitle  \undefined \def \bctitle#1{#1}\fi
\ifx \beditor  \undefined \def \beditor#1{#1}\fi
\ifx \bpublisher  \undefined \def \bpublisher#1{#1}\fi
\ifx \bbtitle  \undefined \def \bbtitle#1{#1}\fi
\ifx \bedition  \undefined \def \bedition#1{#1}\fi
\ifx \bseriesno  \undefined \def \bseriesno#1{#1}\fi
\ifx \blocation  \undefined \def \blocation#1{#1}\fi
\ifx \bsertitle  \undefined \def \bsertitle#1{#1}\fi
\ifx \bsnm \undefined \def \bsnm#1{#1}\fi
\ifx \bsuffix \undefined \def \bsuffix#1{#1}\fi
\ifx \bparticle \undefined \def \bparticle#1{#1}\fi
\ifx \barticle \undefined \def \barticle#1{#1}\fi
\bibcommenthead
\ifx \bconfdate \undefined \def \bconfdate #1{#1}\fi
\ifx \botherref \undefined \def \botherref #1{#1}\fi
\ifx \url \undefined \def \url#1{\textsf{#1}}\fi
\ifx \bchapter \undefined \def \bchapter#1{#1}\fi
\ifx \bbook \undefined \def \bbook#1{#1}\fi
\ifx \bcomment \undefined \def \bcomment#1{#1}\fi
\ifx \oauthor \undefined \def \oauthor#1{#1}\fi
\ifx \citeauthoryear \undefined \def \citeauthoryear#1{#1}\fi
\ifx \endbibitem  \undefined \def \endbibitem {}\fi
\ifx \bconflocation  \undefined \def \bconflocation#1{#1}\fi
\ifx \arxivurl  \undefined \def \arxivurl#1{\textsf{#1}}\fi
\csname PreBibitemsHook\endcsname

%%% 1
\bibitem[\protect\citeauthoryear{Team}{2024}]{team2024eigenlayer}
\begin{botherref}
\oauthor{\bsnm{Team}, \binits{E.}}:
Eigenlayer: The restaking collective.
White paper,
1--19
(2024)
\end{botherref}
\endbibitem

%%% 2
\bibitem[\protect\citeauthoryear{Nofer et~al.}{2017}]{nofer2017blockchain}
\begin{barticle}
\bauthor{\bsnm{Nofer}, \binits{M.}},
\bauthor{\bsnm{Gomber}, \binits{P.}},
\bauthor{\bsnm{Hinz}, \binits{O.}},
\bauthor{\bsnm{Schiereck}, \binits{D.}}:
\batitle{Blockchain}.
\bjtitle{Business \& information systems engineering}
\bvolume{59},
\bfpage{183}--\blpage{187}
(\byear{2017})
\end{barticle}
\endbibitem

%%% 3
\bibitem[\protect\citeauthoryear{Buterin et~al.}{2013}]{buterin2013ethereum}
\begin{barticle}
\bauthor{\bsnm{Buterin}, \binits{V.}}, \betal:
\batitle{Ethereum white paper}.
\bjtitle{GitHub repository}
\bvolume{1}(\bissue{22-23}),
\bfpage{5}--\blpage{7}
(\byear{2013})
\end{barticle}
\endbibitem

%%% 4
\bibitem[\protect\citeauthoryear{Yakovenko}{2018}]{yakovenko2018solana}
\begin{botherref}
\oauthor{\bsnm{Yakovenko}, \binits{A.}}:
Solana: A new architecture for a high performance blockchain v0. 8.13.
Whitepaper
(2018)
\end{botherref}
\endbibitem

%%% 5
\bibitem[\protect\citeauthoryear{Nakamoto}{2008}]{nakamoto2008bitcoin}
\begin{barticle}
\bauthor{\bsnm{Nakamoto}, \binits{S.}}:
\batitle{Bitcoin whitepaper}.
\bjtitle{URL: https://bitcoin. org/bitcoin. pdf-(: 17.07. 2019)}
\bvolume{9},
\bfpage{15}
(\byear{2008})
\end{barticle}
\endbibitem

%%% 6
\bibitem[\protect\citeauthoryear{Monnet}{2011}]{monnet2011rehypothecation}
\begin{botherref}
\oauthor{\bsnm{Monnet}, \binits{C.}}:
Rehypothecation.
Business Review, Federal Reserve Bank of Philadelphia,(Q4),
18--25
(2011)
\end{botherref}
\endbibitem

%%% 7
\bibitem[\protect\citeauthoryear{{Symbiotic Protocol}}{2025}]{symbiotic2025}
\begin{botherref}
\oauthor{\bsnm{{Symbiotic Protocol}}}:
Symbiotic: Permissionless Restaking Protocol.
\url{https://docs.symbiotic.fi/}.
Accessed: 2025-04-23
(2025)
\end{botherref}
\endbibitem

%%% 8
\bibitem[\protect\citeauthoryear{{Babylon Labs}}{2025}]{babylonlabs2025}
\begin{botherref}
\oauthor{\bsnm{{Babylon Labs}}}:
Babylon Labs Documentation.
\url{https://docs.babylonlabs.io/}.
Accessed: 2025-04-23
(2025)
\end{botherref}
\endbibitem

%%% 9
\bibitem[\protect\citeauthoryear{{Jito Labs}}{2025}]{jitolabs2025}
\begin{botherref}
\oauthor{\bsnm{{Jito Labs}}}:
Jito Labs Documentation.
\url{https://docs.jito.wtf/}.
Accessed: 2025-04-23
(2025)
\end{botherref}
\endbibitem

%%% 10
\bibitem[\protect\citeauthoryear{Castro et~al.}{1999}]{castro1999practical}
\begin{bchapter}
\bauthor{\bsnm{Castro}, \binits{M.}},
\bauthor{\bsnm{Liskov}, \binits{B.}}, \betal:
\bctitle{Practical byzantine fault tolerance}.
In: \bbtitle{OsDI},
vol. \bseriesno{99},
pp. \bfpage{173}--\blpage{186}
(\byear{1999})
\end{bchapter}
\endbibitem

%%% 11
\bibitem[\protect\citeauthoryear{Durvasula and Roughgarden}{2024}]{durvasula2024robust}
\begin{botherref}
\oauthor{\bsnm{Durvasula}, \binits{N.}},
\oauthor{\bsnm{Roughgarden}, \binits{T.}}:
Robust restaking networks.
arXiv preprint arXiv:2407.21785
(2024)
\end{botherref}
\endbibitem

%%% 12
\bibitem[\protect\citeauthoryear{Chitra and Pai}{2024}]{chitra2024much}
\begin{botherref}
\oauthor{\bsnm{Chitra}, \binits{T.}},
\oauthor{\bsnm{Pai}, \binits{M.}}:
How much should you pay for restaking security?
arXiv preprint arXiv:2408.00928
(2024)
\end{botherref}
\endbibitem

%%% 13
\bibitem[\protect\citeauthoryear{Markowitz}{2010}]{markowitz2010portfolio}
\begin{barticle}
\bauthor{\bsnm{Markowitz}, \binits{H.M.}}:
\batitle{Portfolio theory: as i still see it}.
\bjtitle{Annu. Rev. Financ. Econ.}
\bvolume{2}(\bissue{1}),
\bfpage{1}--\blpage{23}
(\byear{2010})
\end{barticle}
\endbibitem

%%% 14
\bibitem[\protect\citeauthoryear{Bar-Zur and Eyal}{2025}]{bar2025elastic}
\begin{botherref}
\oauthor{\bsnm{Bar-Zur}, \binits{R.}},
\oauthor{\bsnm{Eyal}, \binits{I.}}:
Elastic restaking networks.
arXiv preprint arXiv:2503.00170
(2025)
\end{botherref}
\endbibitem

%%% 15
\bibitem[\protect\citeauthoryear{Boyd and Vandenberghe}{2004}]{boyd2004convex}
\begin{bbook}
\bauthor{\bsnm{Boyd}, \binits{S.P.}},
\bauthor{\bsnm{Vandenberghe}, \binits{L.}}:
\bbtitle{Convex Optimization}.
\bpublisher{Cambridge university press}, \blocation{???}
(\byear{2004})
\end{bbook}
\endbibitem

%%% 16
\bibitem[\protect\citeauthoryear{Owen}{2013}]{owen2013game}
\begin{bbook}
\bauthor{\bsnm{Owen}, \binits{G.}}:
\bbtitle{Game Theory}.
\bpublisher{Emerald Group Publishing}, \blocation{???}
(\byear{2013})
\end{bbook}
\endbibitem

%%% 17
\bibitem[\protect\citeauthoryear{{Catalysis Network}}{2025}]{catalysis2025}
\begin{botherref}
\oauthor{\bsnm{{Catalysis Network}}}:
Catalysis: Enabling a Future with Thousands of AVSs.
\url{https://blog.catalysis.network/blog/enabling-a-future}.
Accessed: 2025-04-23
(2025)
\end{botherref}
\endbibitem

%%% 18
\bibitem[\protect\citeauthoryear{Deb et~al.}{2024}]{deb2024stakesure}
\begin{botherref}
\oauthor{\bsnm{Deb}, \binits{S.}},
\oauthor{\bsnm{Raynor}, \binits{R.}},
\oauthor{\bsnm{Kannan}, \binits{S.}}:
Stakesure: Proof of stake mechanisms with strong cryptoeconomic safety.
arXiv preprint arXiv:2401.05797
(2024)
\end{botherref}
\endbibitem

%%% 19
\bibitem[\protect\citeauthoryear{Lin}{2016}]{lin2016dirichlet}
\begin{botherref}
\oauthor{\bsnm{Lin}, \binits{J.}}:
On the dirichlet distribution.
Department of Mathematics and Statistics, Queens University
\textbf{40}
(2016)
\end{botherref}
\endbibitem

%%% 20
\bibitem[\protect\citeauthoryear{James}{1980}]{james1980monte}
\begin{barticle}
\bauthor{\bsnm{James}, \binits{F.}}:
\batitle{Monte carlo theory and practice}.
\bjtitle{Reports on progress in Physics}
\bvolume{43}(\bissue{9}),
\bfpage{1145}
(\byear{1980})
\end{barticle}
\endbibitem

%%% 21
\bibitem[\protect\citeauthoryear{Nie}{2022}]{nie2022analysis}
\begin{barticle}
\bauthor{\bsnm{Nie}, \binits{C.-X.}}:
\batitle{Analysis of critical events in the correlation dynamics of cryptocurrency market}.
\bjtitle{Physica A: Statistical Mechanics and its Applications}
\bvolume{586},
\bfpage{126462}
(\byear{2022})
\end{barticle}
\endbibitem

%%% 22
\bibitem[\protect\citeauthoryear{Levy}{1992}]{levy1992stochastic}
\begin{barticle}
\bauthor{\bsnm{Levy}, \binits{H.}}:
\batitle{Stochastic dominance and expected utility: Survey and analysis}.
\bjtitle{Management science}
\bvolume{38}(\bissue{4}),
\bfpage{555}--\blpage{593}
(\byear{1992})
\end{barticle}
\endbibitem

%%% 23
\bibitem[\protect\citeauthoryear{Kreps}{1989}]{kreps1989nash}
\begin{bchapter}
\bauthor{\bsnm{Kreps}, \binits{D.M.}}:
\bctitle{Nash equilibrium}.
In: \bbtitle{Game Theory},
pp. \bfpage{167}--\blpage{177}.
\bpublisher{Springer}, \blocation{???}
(\byear{1989})
\end{bchapter}
\endbibitem

%%% 24
\bibitem[\protect\citeauthoryear{Sun et~al.}{2020}]{sun2020model}
\begin{bchapter}
\bauthor{\bsnm{Sun}, \binits{H.}},
\bauthor{\bsnm{Ruan}, \binits{N.}},
\bauthor{\bsnm{Su}, \binits{C.}}:
\bctitle{How to model the bribery attack: A practical quantification method in blockchain}.
In: \bbtitle{Computer Security--ESORICS 2020: 25th European Symposium on Research in Computer Security, ESORICS 2020, Guildford, UK, September 14--18, 2020, Proceedings, Part II 25},
pp. \bfpage{569}--\blpage{589}
(\byear{2020}).
\bcomment{Springer}
\end{bchapter}
\endbibitem

%%% 25
\bibitem[\protect\citeauthoryear{Alvarez and Nojoumian}{2020}]{alvarez2020comprehensive}
\begin{barticle}
\bauthor{\bsnm{Alvarez}, \binits{R.}},
\bauthor{\bsnm{Nojoumian}, \binits{M.}}:
\batitle{Comprehensive survey on privacy-preserving protocols for sealed-bid auctions}.
\bjtitle{Computers \& Security}
\bvolume{88},
\bfpage{101502}
(\byear{2020})
\end{barticle}
\endbibitem

%%% 26
\bibitem[\protect\citeauthoryear{Adomian and Rach}{1985}]{adomian1985coupled}
\begin{barticle}
\bauthor{\bsnm{Adomian}, \binits{G.}},
\bauthor{\bsnm{Rach}, \binits{R.}}:
\batitle{Coupled differential equations and coupled boundary conditions}.
\bjtitle{Journal of mathematical analysis and applications}
\bvolume{112}(\bissue{1}),
\bfpage{129}--\blpage{135}
(\byear{1985})
\end{barticle}
\endbibitem

%%% 27
\bibitem[\protect\citeauthoryear{Chitra et~al.}{2020}]{gauntlet_compound2020}
\begin{botherref}
\oauthor{\bsnm{Chitra}, \binits{T.}},
\oauthor{\bsnm{Chiang}, \binits{R.}},
\oauthor{\bsnm{Morrow}, \binits{J.}},
\oauthor{\bsnm{Kao}, \binits{H.-T.}}:
An Analysis of the Market Risk to Participants in the Compound Protocol.
\url{https://gauntlet.network/reports/compound}.
Accessed: 2025-04-23
(2020)
\end{botherref}
\endbibitem

%%% 28
\bibitem[\protect\citeauthoryear{Ghosh et~al.}{2024}]{ghosh2024compoundv3economicaudit}
\begin{botherref}
\oauthor{\bsnm{Ghosh}, \binits{R.}},
\oauthor{\bsnm{Gupta}, \binits{S.}},
\oauthor{\bsnm{Datta}, \binits{A.}},
\oauthor{\bsnm{Nag}, \binits{A.}},
\oauthor{\bsnm{Sinha}, \binits{S.}}:
Compound V3 Economic Audit Report
(2024).
\url{https://arxiv.org/abs/2410.04085}
\end{botherref}
\endbibitem

%%% 29
\bibitem[\protect\citeauthoryear{Nag et~al.}{2024}]{nag2024multiagentinfluencediagrams}
\begin{botherref}
\oauthor{\bsnm{Nag}, \binits{A.}},
\oauthor{\bsnm{Gupta}, \binits{S.}},
\oauthor{\bsnm{Sinha}, \binits{S.}},
\oauthor{\bsnm{Datta}, \binits{A.}}:
Multi Agent Influence Diagrams for DeFi Governance
(2024).
\url{https://arxiv.org/abs/2402.15037}
\end{botherref}
\endbibitem

%%% 30
\bibitem[\protect\citeauthoryear{{EigenLayer}}{2025}]{eigenlayerSidecar2025}
\begin{botherref}
\oauthor{\bsnm{{EigenLayer}}}:
EigenLayer Sidecar: Getting Started Guide.
\url{https://sidecar-docs.eigenlayer.xyz/docs/sidecar/running/getting-started}.
Accessed: 2025-04-23
(2025)
\end{botherref}
\endbibitem

%%% 31
\bibitem[\protect\citeauthoryear{Catalano et~al.}{2009}]{catalano2009measuring}
\begin{barticle}
\bauthor{\bsnm{Catalano}, \binits{M.T.}},
\bauthor{\bsnm{Leise}, \binits{T.L.}},
\bauthor{\bsnm{Pfaff}, \binits{T.J.}}:
\batitle{Measuring resource inequality: The gini coefficient}.
\bjtitle{Numeracy}
\bvolume{2}(\bissue{2}),
\bfpage{4}
(\byear{2009})
\end{barticle}
\endbibitem

\end{thebibliography}

\pagebreak

\begin{appendices}
\section{Proportional Reward System}\label{rewardapp}
\begin{figure}
    \centering
    \includegraphics[width=0.5\linewidth]{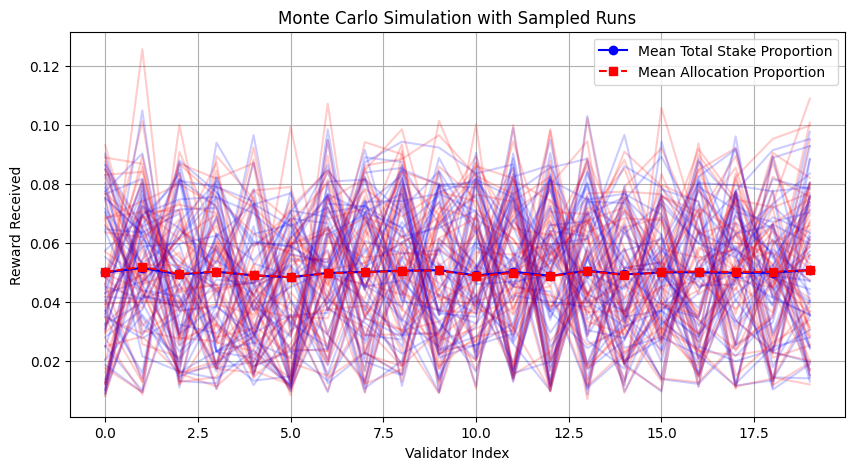}
    \includegraphics[width=0.5\linewidth]{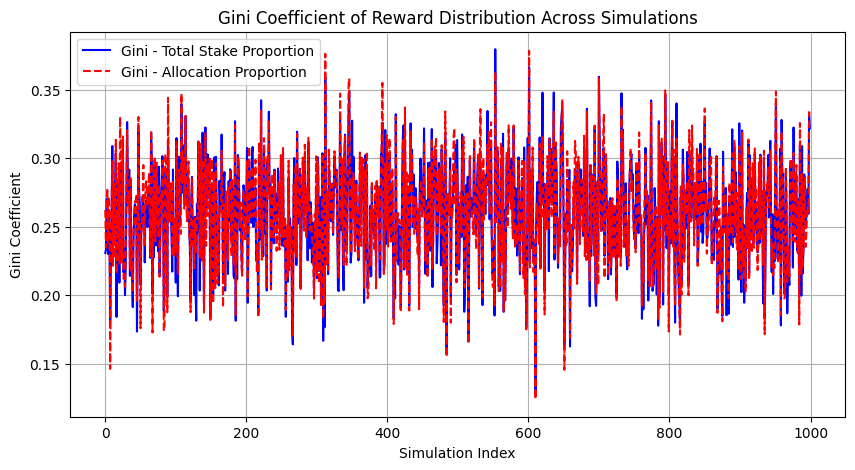}
    \caption{Simulating the reward differences in Total Stake vs Allocated Stake Reward systems - mean is same but higher variations based on different allocation strategies. The Gini index is concentrated around 0.25 which implies relatively even income distribution across node operators when rewards are proportional}
    \label{reward}
\end{figure}

We present two initial simulations to get our initial claim out of the way of having a proportional reward system for each of the node operators in the protocol. We ran 1000 monte carlo simulations of different reward mechanisms and we see that the mean of the rewards is the same whether we divide the rewards based on total stake across services or allocated stake per service however the variation occurs due to the randomization of reward schemes where equality would occur if and only if the reward scheme is proportional (see Figure \ref{reward}). This is proven subsequently where the gini coefficient \cite{catalano2009measuring} is concentrated around $0.25$ in the proportional reward scheme which implies relatively equivalent income distribution.

Assuming a proportional distribution of rewards per SSP as well, this means that 
$$R_i = \frac{\sum_{v} \omega(v,s_i)}{\sum_{s} \sum_{v} \omega(v,s)} \cdot R$$
$$\implies R_i = \frac{\sum_{v} \omega(v,s_i)}{\Delta} \cdot R $$
which implies rewards for any node operator $v_i$ is 
$$u(v_i) = \sum_{j =1}^{k} \frac{\omega(v_i,s_j)}{\sum_{i=1}^{n} \omega(v_i,s_j)} \cdot R_j$$
$$\implies u(v_i) = \sum_{j =1}^{k}\frac{\omega(v_i,s_j)}{\sum_{i=1}^{n} \omega(v_i,s_j)} \cdot \frac{\sum_{i=1}^{n} \omega(v_i,s_j)}{\Delta} \cdot R $$
$$\implies u(v_i) = \sum_{j =1}^{k}\frac{\omega(v_i,s_j)}{\Delta} \cdot R $$
$$\implies u(v_i) = \frac{R}{\Delta} \cdot \sum_{j =1}^{k}\omega(v_i,s_j) = \frac{r \cdot \Delta}{\Delta} \cdot \sum_{j =1}^{k}\omega(v_i,s_j) $$
$$\implies \boxed{u(v_i) = r\cdot \sigma(v_i)}$$

\section{Proofs of Definitions, Theorems and Lemmas}\label{lemproof}

\subsection{Proof of Definition \ref{WES}}
\begin{proof}
For a successfully attackable (i.e., economically insecure) protocol,
\[
\sum_{s} \sum_{v} \alpha(v,s) < \pi.
\]
We also know
\[
\frac{\sum_{v \in V} \alpha(v,s)}{\sum_{v \in V} \omega(v,s)} \geq \theta.
\]
Combining both,
\[
\pi > \sum_{s} \sum_{v} \alpha(v,s) \geq \theta \cdot \sum_{s} \sum_{v} \omega(v,s) = \theta \cdot \Delta.
\]
Hence, for cryptoeconomic security:
\[
\boxed{\pi \leq \theta \cdot \Delta}.
\]
\end{proof}
\subsection{Proof of Definition \ref{SES}}
\begin{proof}
For strong security, each node operator must prefer honesty. That is, for each $v$:
\[
u_v(\alpha) < u(v),
\]
\[
\Rightarrow \gamma(v,\alpha)\cdot \pi - c(v,\alpha) < r\cdot\sigma(v_i),
\]
\[
\Rightarrow \pi < \sum_{j=1}^{k} \alpha(v,s_j) + r\cdot\sigma(v_i),
\]
since $\gamma(v,\alpha) \in [0,1]$.

Summing over all validators:
\[
\pi \cdot n < \sum_{i=1}^{n} \sum_{j=1}^{k} \alpha(v_i,s_j) + r\cdot\sum_{i=1}^{n} \sigma(v_i) = \sum_{i=1}^{n} \sum_{j=1}^{k} \alpha(v_i,s_j) + R.
\]
Thus,
\[
\boxed{\pi < \frac{1}{n} \sum_{i=1}^{n} \sum_{j = 1}^{k} \alpha(v_i,s_j) + \frac{R}{n}}.
\]
\end{proof}

\subsection{Proof of Lemma \ref{lemma:validator-margin}}
\begin{proof}
For each SSP:
\[
\theta \cdot \Delta_j > \sum_{i} \alpha(v_i, s_j).
\]
Summing over $j$:
\[
\theta \cdot \Delta = \sum_{j=1}^k \theta \cdot \Delta_j > \sum_{j=1}^k \sum_{i=1}^n \alpha(v_i, s_j).
\]
Adding $r \cdot \Delta$ to both sides:
\[
(\theta + r) \cdot \Delta > \sum_{j=1}^k \sum_{i=1}^n \alpha(v_i, s_j) + r \cdot \Delta.
\]
Divide by $n$ to obtain the result.
\end{proof}
\subsection{Proof of Lemma \ref{attsur}}
\begin{proof}
In single SSP or $\mathbb{S}$:
\[
\sum_{v \in V} \alpha(v, s_1) \geq \theta \cdot \Delta.
\]
In multi-SSP $\mathbb{M}$:
\[
\exists j: \sum_{v \in V} \alpha(v, s_j) \geq \theta \cdot \Delta_j.
\]
Hence, there are $k$ possible points of failure. Now
Let $\Delta = \sum_j \Delta_j$.
Then:
\[
\min_j \Delta_j < \Delta \Rightarrow \theta \cdot \min_j \Delta_j < \theta \cdot \Delta.
\]
Thus,
\[
C_{\mathbb{M}} < C_{\mathbb{S}}.
\]
\end{proof}

\subsection{Proof of Lemma \ref{stochdom}}
\begin{proof}
First order Stochastic Dominance implies:
\[
\mathbb{P}(\Delta_j(t) \leq x) \geq \mathbb{P}(\Delta_k(t) \leq x), \quad \forall x.
\]
Then for monotonic increasing $f$:
\[
\mathbb{E}[f(\Delta_j(t))] \leq \mathbb{E}[f(\Delta_k(t))].
\]
Taking $f(x) = x$:
\[
\mathbb{E}[\Delta_j(t)] \leq \mathbb{E}[\Delta_k(t)].
\]
Since:
\[
\min_j \Delta_j(t) \leq \Delta_k(t),
\quad \Rightarrow \mathbb{E}[\min_j \Delta_j(t)] \leq \mathbb{E}[\Delta_k(t)].
\]
Multiplying both sides by \( \theta \):
\[
\mathbb{E}[C_{\min}] \leq \theta \cdot \mathbb{E}[\Delta_k(t)].
\]
\end{proof}

\subsection{Proof of Theorem \ref{lem:concavity}}
\begin{proof}
Let \( x, y \in \mathbb{R}_{\geq 0}^k \) and \( \lambda \in [0,1] \). Then,
\[
S(\lambda x + (1 - \lambda)y) = \theta \cdot \min_j \left( \lambda x_j + (1 - \lambda)y_j \right).
\]
By the concavity of the \( \min \) function over \( \mathbb{R}^k \), we have:
\[
\min_j \left( \lambda x_j + (1 - \lambda)y_j \right) \geq \lambda \min_j x_j + (1 - \lambda) \min_j y_j.
\]
Multiplying both sides by \( \theta > 0 \), we obtain:
\[
S(\lambda x + (1 - \lambda)y) \geq \lambda S(x) + (1 - \lambda) S(y).
\]
Hence, \( S(x) \) is concave on \( \mathbb{R}_{\geq 0}^k \).
Now
Let the feasible set be:
\[
\Omega := \left\{ \omega \in \mathbb{R}_+^{n \times k} \; \middle| \; \sum_{j=1}^k \omega(v_i, s_j) = \sigma(v_i), \; \forall i \in [n] \right\}.
\]
Define the objective function:
\[
f(\omega) := \min_{j \in [k]} \Delta_j(\omega) = \min_{j \in [k]} \left( \sum_{i=1}^n \omega(v_i, s_j) \cdot p_j(t) \right).
\]
Each \( \Delta_j(\omega) \) is a linear function of the decision variables \( \omega \), and hence \( f(\omega) \), as the pointwise minimum over affine functions, is concave.

Since the objective is concave and the constraints define a convex set \( \Omega \), the problem is a convex program:
\[
\max_{\omega \in \Omega} f(\omega).
\]
Convex optimization theory tells us that the maximum of \( \min_j \Delta_j \) under linear constraints is achieved when the values \( \Delta_j \) are equalized as much as possible. If \( \Delta_j \neq \Delta_\ell \) for some \( j \neq \ell \), it is always possible to reallocate a small amount of stake from the more secure SSP to the weaker one without violating the constraints, thereby increasing the minimum and improving the objective.

Hence, at optimality:
\[
\Delta_j^* = \Delta_\ell^*, \quad \forall j, \ell \in [k].
\]
\end{proof}
 \subsection{Proof of Lemma \ref{Equil}}
\begin{proof}
    When $\Delta_j = \Delta$ for all $j$, then $\frac{R}{\Delta_j}$ is constant. The validator's utility simplifies to:
    \[
    u(v_i) = \frac{R}{\Delta} \sum_{j} \omega(v_i, s_j) = \frac{R}{\Delta} \cdot \sigma(v_i)
    \]
\end{proof} 
\pagebreak
\section{Table of Notations Used in the Paper}\label{notations}
\begin{table}[h!]
\centering
\caption{Table of Notations}
\begin{tabular}{ll}
\toprule
\textbf{Symbol} & \textbf{Description} \\
\midrule
$V = \{v_1, \ldots, v_n\}$ & Set of $n$ validators (node operators) \\
$S = \{s_1, \ldots, s_k\}$ & Set of $k$ SSPs (Shared Security Providers) \\
$\sigma(v_i)$ & Total stake held by validator $v_i$ \\
$\omega(v_i, s_j)$ & Stake allocated by $v_i$ to SSP $s_j$ \\
$\Delta$ & Total stake across all SSPs ($\sum_i \sigma(v_i)$) \\
$\Delta_j$ & Total stake allocated to SSP $s_j$ \\
$\Omega$ & Feasible allocation set across validators and SSPs \\
$\pi$ & Profit from a successful attack on the AVS \\
$\theta$ & Attack threshold (e.g., $1/3$ in PBFT) \\
$R$ & Total reward distributed by the AVS \\
$r$ & Annual percentage yield (APY) as a fraction of total stake \\
$\alpha(v_i, s_j)$ & Stake used by $v_i$ to attack the AVS via $s_j$ \\
$c(v_i, \alpha)$ & Cost incurred by $v_i$ to attack the AVS \\
$C(\alpha)$ & Total cost of attack across all validators \\
$u(v_i)$ & Utility of validator $v_i$ from honest participation \\
$u_v(\alpha)$ & Utility of $v_i$ from attacking the AVS \\
$p_j(t)$ & Price of the asset restaked in SSP $s_j$ at time $t$ \\
$\Delta_j(t)$ & security level of SSP $s_j$ \\
$\lambda_j(v_i)$ & Per-unit bribe required for $v_i$ in SSP $s_j$ \\
$C_{\text{multi}}, C_{\text{single}}$ & Bribery cost in Model M and Model S respectively \\
$S(x) = \theta \cdot \min_j x_j$ & Security function over SSP allocations \\
$\omega^*$ & Equilibrium or optimal allocation vector \\
$b(v_i,s_j)$ & Bribe given to validator $v_i$ in SSP $s_j$\\

\bottomrule
\end{tabular}
\label{tab:notations}
\end{table}




\end{appendices}

\end{document}